\documentclass[pre,aps,twocolumn,superscriptaddress,showpacs]{revtex4}

\usepackage[dvips]{graphicx}
\usepackage{amssymb,amsfonts,amsmath}
\usepackage{color}
\usepackage{ulem}

\newcommand{\ups}{\Upsilon}
\newcommand{\vt}{\vartheta}
\newcommand{\vp}{\varphi}
\newcommand{\X}{\boldsymbol{\chi}}

\newcommand{\F}{\boldsymbol{F}}

\newcommand{\pp}{\boldsymbol{p}}
\newcommand{\J}{\boldsymbol{J}}

\newcommand{\pvec}{\boldsymbol{p}}
\newcommand{\rr}{\boldsymbol{r}}

\newcommand{\xxi}{\boldsymbol{\xi}}
\newcommand{\xchi}{\boldsymbol{\chi}}
\newcommand{\eeta}{\boldsymbol{\eta}}
\newcommand{\R}{\boldsymbol{R}}

\begin{document}

\title{Effective Interactions in Active Brownian Suspensions}

\author{T.~F.~F. Farage}
\affiliation{Department of Physics,
  University of Fribourg, CH-1700 Fribourg, Switzerland}
\author{P. Krinninger}
\affiliation{Theoretische Physik II, Physikalisches Institut, 
  Universit{\"a}t Bayreuth, D-95440 Bayreuth, Germany}
\author{J.~M. Brader$^1$}

\begin{abstract}
Active colloids 
exhibit persistent motion, which can lead to motility-induced phase separation (MIPS). 
However, there currently exists no microscopic theory to account for this phenomenon.  
%
We report a first-principles theory, free of fit parameters, for active spherical colloids, 
which shows explicitly how an {\it effective} many-body interaction potential is generated 
by activity and how this can rationalize MIPS.  
For a passively repulsive system the theory predicts 
phase separation and pair correlations in quantitative agreement with simulation. 
For an attractive system the theory shows that phase 
separation becomes suppressed by moderate activity, consistent with recent 
experiments and simulations, 
and suggests a mechanism for reentrant cluster formation at high activity. 
\end{abstract}

\maketitle

Active colloidal particles in suspension are currently the 
subject of considerable attention, due largely to their ability to model 
self-organisation phenomena in biological systems, but also as a new branch 
of fundamental research in nonequilibrium statistical mechanics: 
assemblies of active colloids are intrinsically out-of-equilibrium systems. 
In contrast to their passive counterparts, active colloids undergo both solvent 
induced Brownian motion and a self-propulsion which requires a continual consumption of 
energy from the local environment.  
Several idealized experimental model systems have been developed, such as catalytic 
Janus particles \cite{lyderic2010,erbe2008,howse2007}, colloids with artificial 
flagella \cite{flagella2005} and light activated particles \cite{palacci2013}. 
The understanding of active systems has been further aided by the development of 
simple theoretical models, which aim to capture the essential physical mechanisms 
and which have been used to study e.g. bacteria, cells or filaments in the cytoskeleton
\cite{collective_motion2012, ramaswamy2010, romanczuk2012, cates_bacteria2012}.

Active particles are characterised by a persistent motion, which can lead to 
`self trapping' dynamics and a rich variety of related collective phenomena 
\cite{collective_motion2012, ramaswamy2010, romanczuk2012, cates_bacteria2012, 
cates_tailleur2014}. 
Even the simplest models of active spherical particles with purely repulsive 
interactions can display the phenomenon of motility-induced phase separation (MIPS) 
\cite{cates_tailleur2014}.  
In many respects, MIPS resembles the equilibrium phase separation familiar from 
passive systems with an attractive component to the interaction potential 
(e.g.~the Lennard-Jones 
potential) 
\cite{tailleur2008, fily2012, stenhammar2013, redner2013, levis_berthier2014}. 
This apparent similarity has motivated several recent attempts to map 
an assembly of active particles onto a passive equilibrium system, 
interacting via an effective attraction  
(usually taken to be a very short range sticky-sphere potential 
\cite{schwarz-linek2012,bocquetPRX_2014}). 
Despite the intuitive appeal of mapping to an equilibrium system, there 
exists no systematic theoretical approach capable of predicting an effective 
equilibrium potential directly from the bare interactions.	

Our current understanding of MIPS has largely been gained 
through either simulation 
\cite{fily2012,redner2013,stenhammar2013,levis_berthier2014,wysocki2014} 
or phenomenological theory 
\cite{tailleur2008,cates2013,stenhammar2013,cates_tailleur2014}. 
The phenomenological theory is based on an  
equation for the coarse-grained density, featuring a local speed  
and a local orientational relaxation time. 
Although the precise relationship between these one-body fields and the 
interparticle interaction potential remains to be clarified, some progress 
in this direction has been made \cite{bialke2013}.
On a more microscopic level, it has recently been shown that a general 
system of active particles
does not have an equation of state \cite{cates_kardar}, due to the influence of the 
confining boundaries, however, one can be recovered for the special case 
of active Brownian spheres \cite{cates_kardar,takatori}.


Here we report a first-principles theory for systems of active Brownian 
spheres, which demonstrates explicitly how an effective 
many-body interaction potential is induced by activity. 
An appealing feature of this approach is that intuition gained from equilibrium can be 
used to understand the steady-state properties of active systems.  
The required input quantities are the passive (`bare') interaction potential, 
the rotational diffusion coefficient and the particle propulsion speed. 
The theory generates as output the static correlation functions and 
phase behaviour of the active system. 
For a repulsive bare interaction, activity generates an attractive 
effective pair potential, thus providing an intuitive explanation for the 
MIPS observed in simulations \cite{fily2012,levis_berthier2014,stenhammar2014}. 
For an attractive bare potential, we find that increasing activity 
first reduces the effective attraction, consistent with the experiments of 
Schwarz-Linek {\it et al.} \cite{schwarz-linek2012}, before leading at higher activity 
to the development of a repulsive potential barrier. 
We speculate that this barrier may be related to the reentrant phase behaviour 
observed in simulation by Redner {\it et al.} \cite{redner2013}. 


The paper will be structured as follows: 
In \S\ref{theory} we specify the microscopic dynamics and 
describe how to eliminate orientational degrees of freedom. 
From the resulting coarse-grained, non-Markovian Langevin equation we 
derive a Fokker-Planck equation for the positional degrees of freedom, 
from which we identify an effective pair potential. 
In \S\ref{results} we employ the effective pair potential in an equilibrium 
integral equation theory and investigate the structure and phase behaviour of 
both repulsive and 
attractive bare potentials. In the former case we predict MIPS, whereas 
in the latter case phase separation is suppressed by activity. 
Finally, in \S\ref{discussion} we discuss our findings and provide an outlook 
for future research.

\section{Theory}\label{theory}
\subsection{Microscopic dynamics}  
We consider a three dimensional system of $N$ active, interacting, spherical Brownian 
particles with spatial coordinate 
$\rr_i$ and orientation specified by an embedded unit vector $\pvec_i$. 
Each particle experiences a self propulsion of speed $v_0$ in its direction of orientation.
Omitting hydrodynamic interactions the particle motion can be modelled by the 
overdamped Langevin equations
\begin{align}
&\dot{\rr}_i = v_0\,\pvec_i  + \gamma^{-1}\F_i + \xxi_i, 
\label{full_langevin}
\\
&\dot{\pvec}_i = \eeta_i\times\pvec_i ,
\label{langevin_orientation}
\end{align}
where $\gamma$ is the friction coefficient and the force on particle $i$ is generated from the 
total potential energy according to $\F_i\!=\!-\nabla_i U_N$.
The stochastic vectors $\xxi_i(t)$ and $\eeta_i(t)$ are Gaussian distributed with zero mean and 
have time correlations	
$\langle\xxi_i(t)\xxi_j(t')\rangle=2D_t\boldsymbol{1}\delta_{ij}\delta(t-t')$ and 
$\langle\eeta_i(t)\eeta_j(t')\rangle=2D_r\boldsymbol{1}\delta_{ij}\delta(t-t')$, where 
$D_t$ and $D_r$ are the translational and rotational diffusion coefficients. 

Equations \eqref{full_langevin} and \eqref{langevin_orientation} are convenient for simulation, 
but are perhaps not the most suitable starting point for developing a first-principles microscopic 
theory. 
For a homogeneous system, averaging over the angular degrees of freedom generates a 
coarse-grained equation \cite{fily2012} 
\begin{align}\label{effective_langevin}
\dot{\rr}_i(t) &= \gamma^{-1}\F_i(t) + \xxi_i(t) + \xchi_i(t), 
\end{align}
where $\xchi_i(t)$ is a Markov process with zero mean and where the
time correlation function is given by
\begin{align}\label{exp_correlation}
\langle\xchi_i(t)\,\xchi_j(t')\rangle = \frac{v_0^2}{3} e^{-2D_r|t-t'|}\boldsymbol{1}\delta_{ij}.   
\end{align} 
The average in \eqref{exp_correlation} is over both noise and initial orientation. 
The distribution of $\xchi_i(t)$ is Gaussian to a good approximation.
This point and further technical details of the coarse graining are discussed 
in Appendix \ref{appendix1}. 
Equation \eqref{effective_langevin} provides a mean-field level of description, which 
deviates from the exact equations \eqref{full_langevin} and \eqref{langevin_orientation} 
by neglecting the coupling of fluctuations in orientation and positional degrees 
of freedom. 

The Langevin equation \eqref{effective_langevin} describes a non-Markovian 
process, which approximates the stochastic time evolution of the 
positional degrees of freedom.  
The persistent motion of active particles is here encoded by 
the exponential decay of the time correlation \eqref{exp_correlation}, with persistence 
time $\tau_p\!=\!(2D_r)^{-1}$\!. 
For small $\tau_p$ the time correlation becomes 
$\langle\xchi_i(t)\xchi_j(t')\rangle = 2D_a\boldsymbol{1}\delta_{ij}\delta(t-t')$ and 
the dynamics reduce to that of an equilibrium system 
with diffusion coefficient $D_t+D_a$, where $D_a\!=\!v_0^2/(6D_r)$. 
This limit is realized when $\tau_p$ is shorter 
than the mean free time between collisions, i.e. in a dilute suspension.  
To treat finite densities requires an approach which deals with 
persistent trajectories. 
With this aim, we adopt \eqref{effective_langevin} as the starting point for contructing  
a closed theory. 

\subsection{Fokker-Planck equation}
A stochastic process driven by colored noise, such as that described by equation 
\eqref{effective_langevin}, is always non-Markovian. 
Consequently, it is not possible to derive an {\it exact} Fokker-Planck equation 
for the time evolution of the probability distribution \cite{vankampen1976}.  
Nevertheless, an approximate Fokker-Planck description capable of making accurate predictions 
can usually be found. 
The approximate Fokker-Planck equation implicitly defines a Markov
process which best approximates the process of physical interest (although precisely what 
constitutes the `best' approximation remains a matter of debate). 
From the extensive literature on this subject (see \cite{vankampen1976,vankampen1998,faetti1988} 
and references therein) has emerged a powerful method due to Fox \cite{fox1986,fox_weak1986}, 
in which a perturbative expansion in powers of correlation time is partially resummed using 
functional calculus. 
The resulting Fokker-Planck equation is most accurate for short correlation times 
(`off white' noise \cite{vankampen1998}) and for one-dimensional models 
makes predictions in good agreement with simulation data \cite{faetti1988}.

We now consider applying the method of Fox \cite{fox1986,fox_weak1986} to 
equation \eqref{effective_langevin}. 
This approach consists of first formulating the configurational probability distribution as 
a path (functional) integral and then making a time-local, Markovian approximation to this quantity. 
Technical details of the method are given in Appendix \ref{appendix2}.
Fox's approach was originally developed to treat one-dimensional problems 
\cite{fox1986,fox_weak1986}, however the generalization to three dimensions is 
quite straightforward.  
This enables us to directly obtain the following Fokker-Planck equation  
\begin{align}
	\partial_t\Psi(\rr^N\!\!,t) = -\sum_{i=1}^{N}\nabla_i\cdot \J_i(\rr^N\!\!,t) ,
\label{smol_eq}
\end{align}
where $\Psi(\rr^N\!\!,t)$ is the configurational probability distribution.  
Within the generalized Fox approximation the many-body current is given by
\begin{align}\label{current_i}
	\J_i(\rr^N\!\!,t) = -D_i(\rr^N)\left[\,\nabla_i 
-\beta \F_i^{\rm eff}(\rr^N)\,\right] \Psi(\rr^N\!\!,t) ,
\end{align}
where $\beta\!\equiv\!(k_BT)^{-1}$. 
The diffusion coefficient is given by 
\begin{align}\label{diffusion_i}
D_i(\rr^N) = D_t + D_a\left(
1 +
\frac{\tau \nabla_i\!\cdot\!\beta\F_i(\rr^N)}{1 - \tau \nabla_i\!\cdot\!\beta\F_i(\rr^N)}
\right), 
\end{align}
where we have defined a dimensionless persistence time, $\tau\!=\!\tau_p D_t/d^2$.  
The effective force is given by
\begin{align}
\F_i^{\rm eff}(\rr^N) =& 
 \frac{1}{\mathcal{D}_i(\rr^N)}\Big( 
\F_i(\rr^N) - k_BT\,\nabla_i \mathcal{D}_i(\rr^N)
\Big),
\label{force_i}
\end{align}
where $\mathcal{D}_i(\rr^N)\!=\!D_i(\rr^N)/D_t$ is a dimensionless diffusion coefficient. 
Either in the absence of interactions or in limit of large $D_r$ 
the diffusivity \eqref{diffusion_i} reduces to
$D_t\!+\!D_a$ and the effective force becomes 
$D_t \F_i(\rr^N)/(D_t+D_a)$. 
In this diffusion limit the system behaves as an equilibrium system at 
effective temperature $T_{\rm eff}=T(1+D_a/D_t)$. 

For weakly persistent motion, $\tau\!\rightarrow\! 0$, equations 
\eqref{smol_eq} to \eqref{force_i} become exact and the theory provides 
the leading order correction to the diffusion approximation. 
However, the Fox approximation goes beyond this by including contributions 
to all orders in $\tau$. Indeed, detailed studies of one-dimensional systems 
have demonstrated 
good results over a large range of $\tau$ values \cite{faetti1988}. 
The only caveat is that the condition $1\!-\!\tau\, \nabla_i\!\cdot\!\beta\F_i\!>\!0$  
must be satisfied \cite{fox1986,fox_weak1986}. The range of accessible $\tau$ 
values thus depends upon the specific form of the bare interaction potential.

Within our stochastic calculus approach, the effective many-body force 
\eqref{force_i} emerges in a natural way from the coarse grained Langevin equation 
\eqref{effective_langevin}.
The more standard route (adopted in all attempts made so 
far \cite{pototsky2012,bialke2013}) to approach this problem is to derive from the 
Markovian equations \eqref{full_langevin} the exact Fokker-Planck 
equation for the joint distribution of positions and orientations, $P(\rr^N\!\!,\pvec^N\!\!,t)$. 
However, coarse graining strategies based on integration of $P$ over 
orientations generate intractable integral terms.  
By starting from \eqref{effective_langevin} we are able to circumvent these difficulties. 
As we shall demonstrate below, our effective force accounts for several important 
collective phenomena in active systems.

\subsection{Effective pair potential}
In the low density limit we need only consider isolated pairs of particles. 
In this limit \eqref{smol_eq} reduces to an equation of motion for the radial 
distribution function, $g(r,t)\!\equiv\!\Psi(r,t)/\rho_b^2$, where 
$\rho_b$ is the bulk density. 
This equation of motion, the pair Smolochowski equation, is given by
\begin{align}
\partial_tg(r,t)\!=\!-\nabla\cdot {\bf j}(r,t), 
\end{align}
where $r\!=\!|\rr_{12}|$ is the particle separation and $\nabla\!=\!\nabla_{\rr_{\!12}}$.   
The pair current is given by
\begin{align}\label{current_pair}
	{\bf j}(r,t) = -2 D(r) g(r,t)\left[\,\nabla\ln g(r,t)
- \beta \F^{\rm eff}(r)\,\right].
\end{align}
where the radial diffusivity 
\begin{align}\label{diffusion_pair}
D(r) = D_t + D_a\left(
1 -
\frac{\tau \nabla^2\beta u(r)}{1 + \tau \nabla^2 \beta u(r)}
\right), 
\end{align}
interpolates between the value $D_t$ at small separations, where $u(r)$ is strongly repulsive, 
and $D_t+D_a$ at large separations. 
The effective interparticle force is given by
\begin{align}
\F^{\rm eff}(r) = 
 \frac{1}{\mathcal{D}(r)}\Big( 
\F(r) - k_BT\,\nabla\mathcal{D}(r)
\Big),
\label{force_pair}
\end{align}
where the bare force is related to the pair potential by 
$\F(r)\!=\!-\nabla u(r)$. 
The symmetry of the two-body problem can be exploited to calculate from 
\eqref{force_pair} an effective interaction 
potential	 
\begin{align}
\beta u^{\rm eff}(r)=\int_r^{\infty}\!\!dr' \left( \frac{\beta F(r')}{\mathcal{D}(r')} 
- \frac{\partial}{\partial r'}\ln \mathcal{D}(r') \right),  
\label{potential_pair}
\end{align}
where $F(r)\!=\!|\F(r)|$. 
We have thus identified an effective interaction pair potential, which requires as input 
the bare potential and the activity parameters $\tau$ and $D_a$. 

\begin{figure}[t!]
\hspace*{-0.1cm}\includegraphics[width=8.65cm]{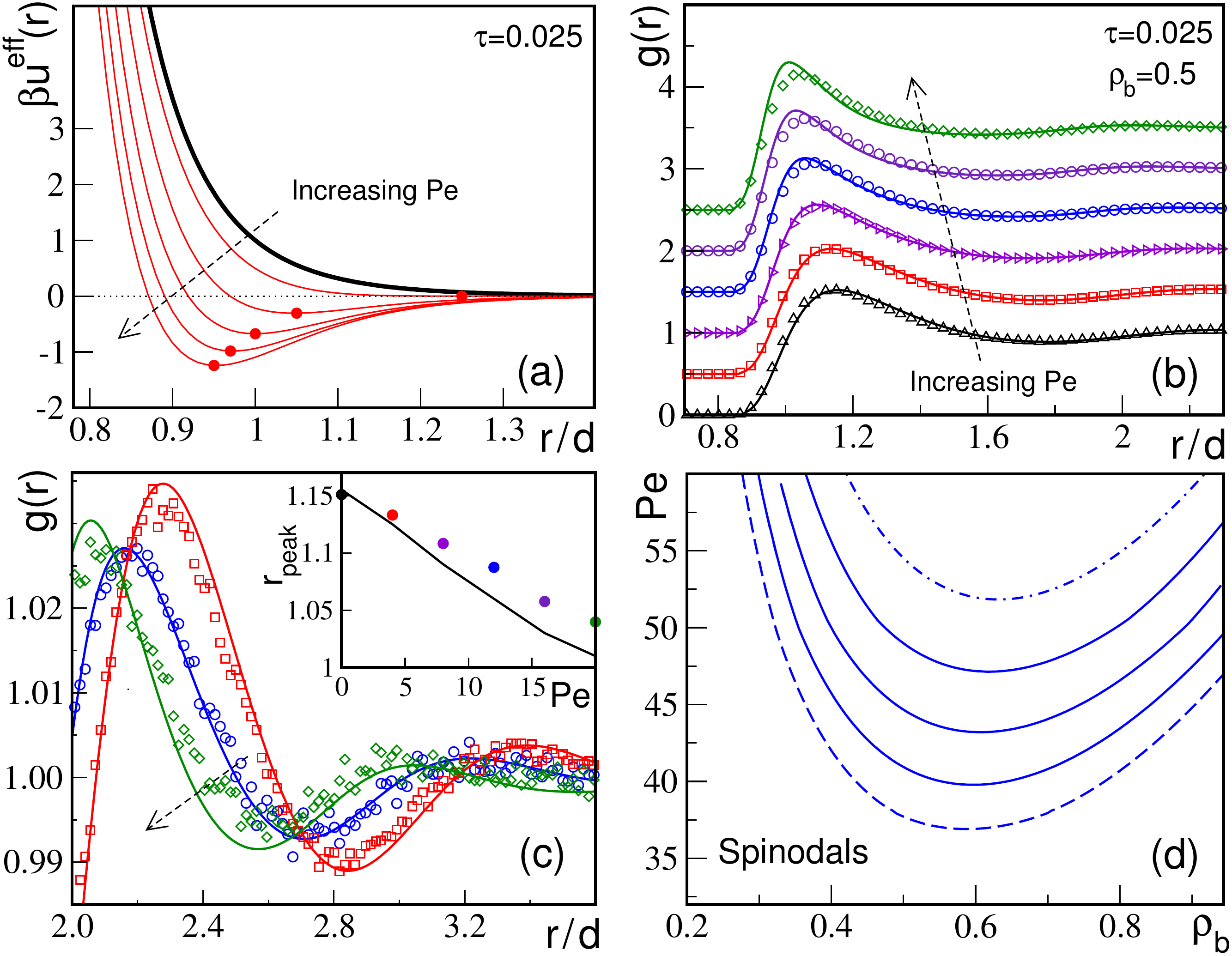}
\vspace*{0.2cm}
\caption{\label{r12_fig}{\bf Activity induces effective attraction.}
Passive potential $\beta u(r)\!=\!r^{-12}\!$.
{\bf(a)} Increasing $Pe$ (in steps of $8$) from $0$ to $40$ 
generates an effective interparticle attraction. 
Points indicate the potential minima. 
{\bf(b)} Radial distribution function, $g(r)$, from simulation (points) and theory (lines) 
for $\rho_b\!=\!0.5$ and $Pe\!=\!0$ to $20$ (in steps of $4$). Curves are shifted vertically 
for clarity.  
{\bf(c)} As in (b), but focusing on larger separations for $Pe=4$ (squares), 
$12$ (circles) and $20$ (diamonds). Inset: position of the first peak in $g(r)$ as a function 
of $Pe$. 
{\bf(d)} Spinodals for $\tau=0.045$ (dot-dashed) to $0.065$ (long dashed) in steps of 
$0.005$.   
} 
\end{figure}

\section{Results}\label{results}
\subsection{Motility-induced phase separation (MIPS)}
To illustrate how activity can generate an effective attraction in a passively repulsive system 
we consider the non-specific potential $\beta u(r)=r^{-12}$. 
In Fig.\ref{r12_fig}a we show the evolution of the effective potential \eqref{potential_pair} 
for fixed $\tau$ as a function of the dimensionless velocity $Pe\!=\!v_0d/D_t$. 
For $Pe\!\gtrsim\!10$ the effective potential develops an attractive tail. 
As $Pe$ is increased the potential well deepens, the minimum moves to 
smaller separations and the radius of the soft repulsive core decreases. 
These trends are consistent with the intuitive picture that persistent motion drives soft 
particles into one another (the soft core radius reduces) and 
that they remain dynamically coupled (`trapped') for longer than in the corresponding 
passive system. 
Within our equilibrium picture the trapping is accounted for by the effective attraction. 

For systems at finite density the pair potential \eqref{potential_pair} is an approximation, because 
three- and higher-body interactions will play a role (see equation \eqref{force_i}). 
However, for simplicity we henceforth employ the pair potential \eqref{potential_pair} for 
all calculations, as we anticipate that this will provide the dominant contribution. 
Although corrections to this assumption can be made, they obscure the physical picture and 
come at the expense of a more complicated theory. 
The validity of the pair potential approximation is justified {\it a posteriori} by 
the comparison with simulation for the finite density pair correlations.

In Fig.\ref{r12_fig}b we show the steady-state (isotropic) radial distribution 
function for $\rho_b\!=\!0.5$ for various values of $Pe$.
We employ the effective pair potential \eqref{potential_pair} together with 
liquid state integral equation theory and compare theoretical predictions
with direct Brownian dynamics simulation of equations \eqref{full_langevin} and 
\eqref{langevin_orientation}. 
The integral equation theory we employ is the soft mean-spherical approximation (SMSA) 
proposed by Madden and Rice \cite{madden1980}.   
This approximate closure of the Ornstein-Zernike equation is known to provide reliable 
results for the pair stucture of Lennard-Jones type potentials.
Given the form of the effective pair potential shown in Fig.\ref{r12_fig} the SMSA 
would seem to be a reasonable choice of closure. 
Details of the integral equation theory and the simulation procedure are given in Appendices 
\ref{integralequation} and \ref{simulations}, respectively.  

\begin{figure}[t!]
\hspace*{-0.1cm}\includegraphics[width=8.65cm]{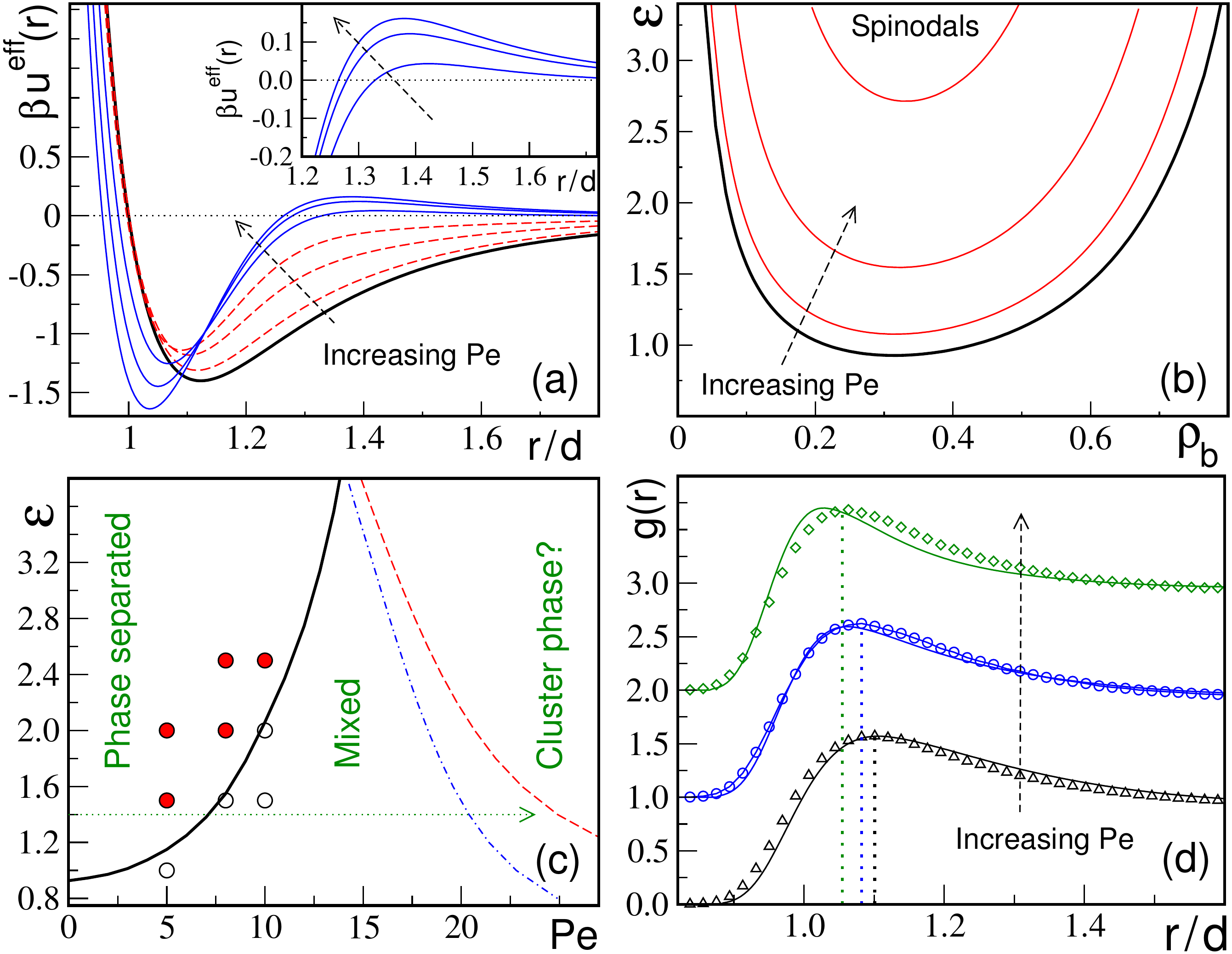}
\vspace*{0.2cm}
\caption{\label{figure2}{\bf From active suppression of phase separation to a cluster phase.} 
Passive potential $\beta u(r)\!=\!4\,\varepsilon(r^{-12}\!-r^{-6})$ with $\varepsilon\!=\!1.4$. 
{\bf (a)} Effective potential \eqref{potential_pair} for $\tau\!=\!0.025$ and 
$Pe\!=\!0$ (black) $4, 8, 12$ (red broken lines) and $20, 28, 36$ (blue).
Inset: zoom of the repulsive peak for $Pe\!=\!20, 28, 36$.
{\bf (b)} Spinodals for $Pe\!=\!0, 4, 8, 12$. Increasing $Pe$ increases $\varepsilon_{\rm crit}$, 
the critical value of $\varepsilon$. 
{\bf (c)} $\varepsilon_{\rm crit}$ as a function of $Pe$ (black full line) 
and the locus of points for which the repulsive peak of $\beta u^{\rm eff}$ 
takes the value $0.1$ (red dashed) and $0.05$ (blue dot-dashed). 
Open (closed) circles indicate points where BD simulation find a mixed (phase-separated) 
state (see figure \ref{sim_snapshot}).
Arrow indicates path taken in (a). 
{\bf (d)} Theory (lines) and simulation (symbols) data (shifted for clarity) 
for $g(r)$ at $\varepsilon=0.5$, $\rho_b=0.3$ for $Pe=0$ (black), $12$ (blue) and $20$ (green). 
Dotted lines indicate peak positions.   
} 
\end{figure}

We find that as $Pe$ is increased the main peak of $g(r)$ grows in height and 
shifts to smaller separations (see 
inset to Fig.\ref{r12_fig}c), reflecting the changes in the effective potential. 
In the main panel of Fig.\ref{r12_fig}c we focus on the second and third peaks. 
The quantitative accuracy of the theory in describing the decay of $g(r)$ 
is quite striking, in particular the phase shift induced by increasing activity 
is very well described. 
Further comparison for other parameter values (not shown) suggests that 
\eqref{potential_pair} combined with the SMSA theory 
provides an accurate account of the asymptotic decay of pair correlations.    

In Fig.\ref{r12_fig}d we show the spinodal lines mapping the locus of points for which the 
static structure factor, $S(q)=(1-\rho_b c(q))^{-1}$, diverges at vanishing wavevector. 
Simulations have shown that MIPS 
is consistent with a spinodal instability \cite{fily2012}. 
As $\tau$ is decreased the critical point moves to higher values of $Pe$ and to slightly higher densities. 
When compared with the spinodal of a standard Lennard-Jones system (e.g. the 
black curve in Fig.\ref{figure2}b) 
the critical points in Fig.\ref{r12_fig}d lie at rather higher 
values of $\rho_b$. 
This suggests that typical coexisting liquid densities for MIPS will be larger than those found  
in equilibrium phase separated systems, as has been observed in simulation 
\cite{fily2012,redner2013}.

\subsection{Suppression of phase separation} 
We next consider the influence of activity on a Lennard-Jones system, 
$\beta u(r)\!=\!4 \varepsilon (r^{-12}\!-r^{-6})$. 
For a phase-separated passive system, recent experiments  
and simulations have demonstrated that increasing $Pe$ first suppresses 
the phase separation \cite{schwarz-linek2012} and then leads at higher $Pe$ to a reentrant MIPS 
\cite{redner2013}. 
Schwarz-Linek {\it et al.} have argued that the suppression of phase separation at lower to intermediate 
$Pe$ occurs in their system because particle pairs bound by the attractive (depletion) 
potential begin to actively escape the potential well, and that this can be mimicked 
using an effective potential less attractive and shorter-ranged than the bare 
potential \cite{schwarz-linek2012}. 

To investigate these phenomena we set $\varepsilon\!=\!1.4$, which ensures a phase separated 
passive state \cite{barker_henderson1976}, and consider the evolution of the 
effective potential as a function of $Pe$. 
In Fig.\ref{figure2}a we show that as $Pe$ is increased from zero to the value $18$ 
both the depth and range 
of the effective potential reduce significantly, consistent with the expectation of 
Schwarz-Linek {\it et al.} 
\cite{schwarz-linek2012}. 
Spinodals within this range of $Pe$ values, identifying where the static structure factor diverges 
at zero wavevector, are shown in Fig.\ref{figure2}b. 
As $Pe$ is increased the critical point moves to higher values of $\varepsilon$ 
(cf. Figs.1 and 3 in Ref.\cite{schwarz-linek2012}). 
A passively phase-separated system will thus revert to a single phase upon increasing the activity. 
To examine this behaviour in more detail we show in Fig.\ref{figure2}c the trajectory of the 
critical point in the ($Pe,\varepsilon$) plane. 
Above the line there bulk densities for which phase separation occurs. 

\begin{figure}[!t]
\hspace*{0.2cm}\includegraphics[width=8.65cm]{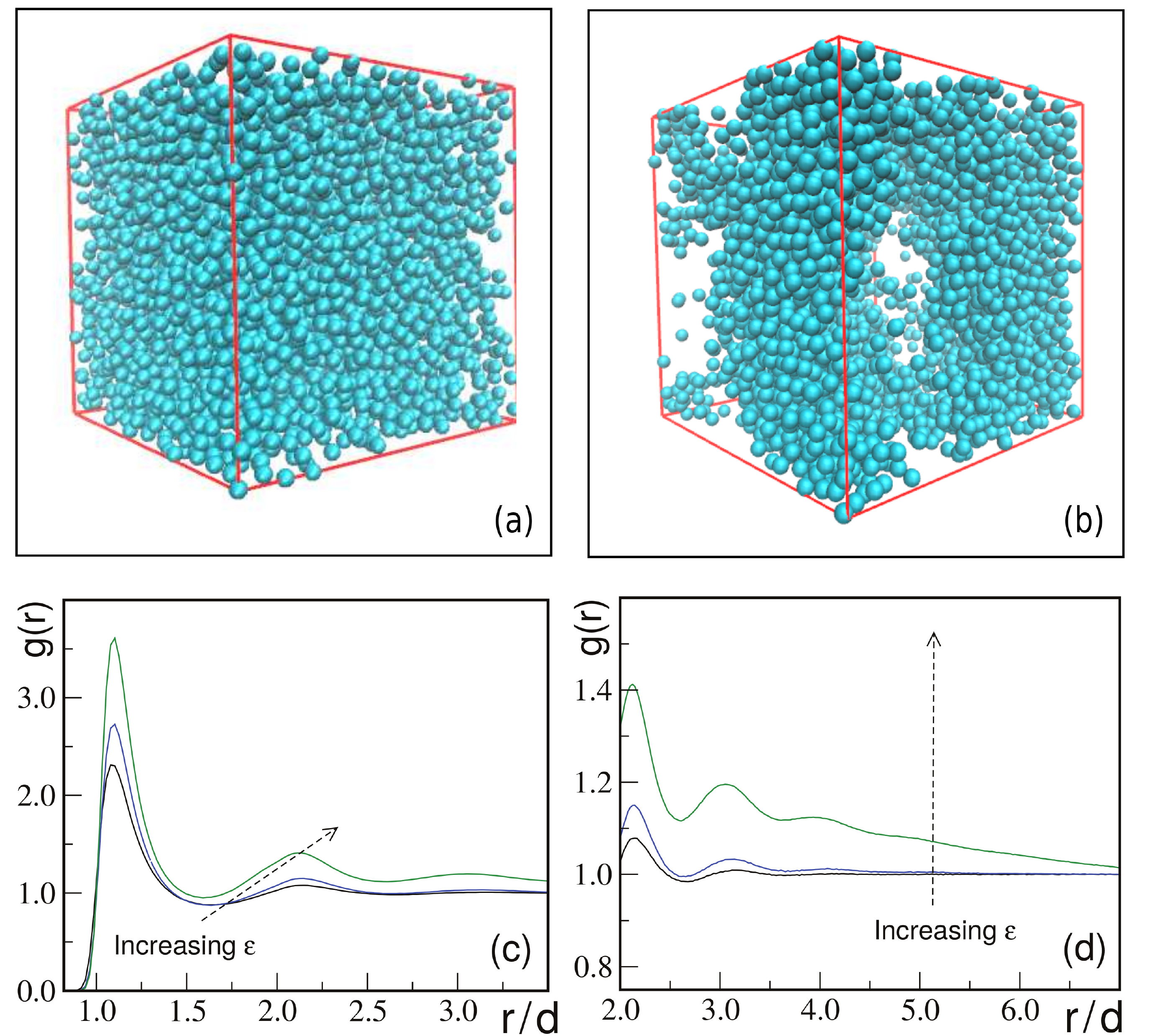}
\vspace*{-0.2cm}
\caption{{\bf Simulated phase separation}. {\bf (a)} Snapshot of a mixed system at 
$t/\tau_B = 40$, $Pe = 8$, $\varepsilon /(k_B T)=1.5$. {\bf (b)} Snapshot of a phase 
separating system at the same time and $Pe=8$, $\varepsilon=2.5$. {\bf (c)} The radial distribution function, 
$g(r)$, for $\varepsilon=1.5$ (black curve), 2 (blue curve), and 2.5 (green 
curve). {\bf (d)} As in (c) but focusing on larger distances. From the 
snapshots together with the long range behaviour of the $g(r)$ we can 
distinguish between a mixed and a phase separated system. The slow decay to the asymptotic value of unity, as shown in (d), indicates phase separation.   
} 
\label{sim_snapshot}
\end{figure}

In order to test the predicted trajectory of the critical point we have performed Brownian 
dynamics simulations at a bulk density $\rho_b\!=\!0.4$, which lies close to the critical 
density \cite{barker_henderson1976}, 
for various values of $\varepsilon$ and the $Pe$ values $5, 8$ and $10$. 
Visual inspection of the simulation snapshots reveals the existence of voids in the particle 
configurations corresponding to a phase separated state 
(see Figs.\ref{sim_snapshot}a and \ref{sim_snapshot}b for snapshots). 
This visual impression can be made more quantitative by calculating the radial 
distribution function. Phase separating states generate a very characteristic 
slow decay of $g(r)$ (Figs.\ref{sim_snapshot}c and \ref{sim_snapshot}d) which 
provides a useful indicator. 
The open circles in Fig.\ref{figure2}c represent mixed states, whereas closed circles 
indicate phase separated statepoints. 
The phase boundary predicted by the theory is highly consistent with the simulation data. 

\subsection{Cluster phase}    
Returning to Fig.\ref{figure2}a, we find that for $Pe\!>\!18$ the effective 
potential develops a repulsive barrier, which grows in height (see inset) with increasing 
$Pe$, while the potential minimum becomes deeper. 
It is well-known that potentials with a short-ranged attraction and long-ranged 
repulsion (SALR potentials) exhibit unusual equilibrium phase behaviour, including clustering 
and microphase-separation \cite{sear1997,archer2007}. 
Although the attractive component of the potential may favour phase separation, the long range repulsion 
destabilises distinct liquid and gas phases and causes them to break up into droplets or clusters. 
This represents a non-spinodal type of phase transition, characterized by a divergence 
in the structure factor at finite wavevector. 
The appearance of a repulsive barrier in the effective potential suggests that a similar 
mechanism may be at work in passively attractive systems subject to high $Pe$ activity.

In Fig.\ref{figure2}c we show the locus of points where the effective potential peak height attains 
a given value (we choose $0.05$ and $0.1$ for illustration). 
When these `isorepulsion curves' are viewed together with the critical point 
trajectory the resulting phase diagram is very similar to that obtained by Redner 
{\it et al.} in their simulation study of two-dimensional active Lennard-Jones particles 
(cf. figure 1 in Ref.\cite{redner2013}). 
However, a detailed study of the connection between the potential barrier and 
high $Pe$ clustering goes beyond the scope of the present work. 


\section{Discussion}\label{discussion}  
In summary, we have shown that systems of active spherical Brownian particles 
can be mapped onto an equilibrium system interacting 
via an effective, activity-dependent many-body potential. 
The only required inputs are the bare potential, thermodynamic statepoint 
and the parameters specifying the state of activity. 
Our theory captures the phenomenon of MIPS in repulsive systems and 
provides first-principles predictions for the activity 
dependence of the pair correlations, in very good agreement with 
Brownian dynamics simulation. 
As far as we are aware there exists no other approach capable of predicting 
from the microscopic interactions the pair correlation functions of an active system. 
Further insight into the steady state particle distribution could in principle be obtained 
by investigating the three-body correlations. 
These could be obtained by employing the effective 
potential in a higher-order liquid state integral equation theory 
(see e.g.~\cite{brader_triplet} and references therein).

For passively attractive systems the theory rationalizes the experimental 
finding \cite{schwarz-linek2012} that increasing activity can suppress passive phase separation.  
We find that as $Pe$ is increased from zero to intermediate values the minimum of the effective 
potential becomes less deep, thus weakening the cohesion of the liquid phase. 
To the best of our  knowledge there is currently no alternative theoretical explanation 
of phase-transition suppression in active suspensions. 
It is an appealing aspect of our theory that the suppression of passive phase separation 
follows naturally from the same approach which yields activity-induced attraction for 
repulsive potentials. 
For high values of $Pe$ the appearance of a repulsive barrier in the effective 
potential suggests that the reentrant phase separation observed in simulations 
\cite{redner2013} may be interpreted using concepts of equilibrium 
clustering in SALR potential systems. 
This will be a subject of future detailed investigations. It is known that 
care must be exercised when analysing SALR potentials, as 
traditional liquid state theories can prove misleading \cite{archer2007}.

%
%

A key step in our development is the Fox approximation \cite{fox1986}, which yields an 
effective Markovian description of the coarse-grained equation \eqref{effective_langevin}.  
Making a Markovian approximation automatically imposes an effective equilibrium,  
however, we are aware that there exist certain situations for which this breaks down 
\cite{cates2013,cates_tailleur2014}. 
Establishing more clearly the range of validity of our approach, as well as its possible 
extensions, will be the subject 
of ongoing study. However, it is already clear that going beyond the Markovian approximation 
will be very challenging. 
Indeed, such a step may not even be desirable. 
Any kind of non-Markovian description 
would lead inevitably to a loss of the effective equilibrium picture and the physical 
intuition associated with it. 
It thus seems likely that 
practical improvements to the present approach will retain the Markovian 
description while seeking to optimize, or improve upon, the Fox approximation for certain 
classes of bare potential. Very recently, Maggi \textit{et al.} have employed 
an alternative approach to treating stochastic processes driven 
by Ornstein-Uhlenbeck noise \cite{maggi2015}. A comparison of their approach 
with the Fox method employed here would be very interesting.

With a view to further applications of our approach, we note that there has 
recently been considerable interest in active suspensions at very high 
densities 
\cite{faragebrader_arxiv,ni2013,berthier2013,szamel_berthier2015}. 
In particular, it has been found using computer simulations that activity has a 
strong influence on the location of the hard-sphere glass 
transition, dynamic correlation functions, such as the intermediate scattering 
function, and static pair correlations \cite{ni2013}. 
Within our effective equilibrium framework, increasing the activity of a passively 
repulsive system generates an effective attraction. 
We can therefore anticipate that for volume fractions just above the glass transition 
it will be possible to observe a reentrant glass transition, namely a melting of 
the glass followed by revitrification, as a function of increasing $Pe$. 
Moreover, the nontrivial evolution of the effective potential as a function of 
$Pe$ for attractive bare potentials (cf.~Fig.\ref{figure2}a) 
suggests these systems will present a rich variety of glassy states. 
Work along these lines is in progress.

Finally, we mention that a natural generalization of the present theory is to 
treat spatially inhomogeneous systems in external fields.  
Recent microscopic studies of active particles under confinement 
(e.g.~in a harmonic trap \cite{pototsky2012}) have provided considerable insight, 
however none of the existing approaches have considered effective {\it interparticle} 
interactions. 
Inhomogeneous generalization of the present theory enables the interaction between 
MIPS and external fields to be investigated on the microscopic level. 
Our preliminary investigations reveal, for example, activity-induced wetting at 
a planar substrate and capillary-condensation under confinement. 
This will be presented in a future publication.

\begin{acknowledgments}
We thank Yaouen Fily, Ronald Fox and Paolo Grigolini for helpful correspondence. 
Funding provided by the Swiss National Science Foundation. 
PK thanks the Elitenetzwerk Bayern (ENB) and Matthias Schmidt for 
financial support.  
\end{acknowledgments}

\appendix
\section{Coarse-grained Langevin equation}\label{appendix1}
Equation \eqref{langevin_orientation} describes the orientational diffusion 
of an active particle. 
%
The corresponding conditional probability distribution function $\ups(\pp,t\,|\,\pp_0, 
t_0)$, where $t>t_0$, obeys a Fokker-Planck equation which can be obtained by 
usual techniques \cite{gardiner},
\begin{equation}\label{FP_p}
 \frac{\partial}{\partial t}\ups(\pp,t\,|\,\pp_0, t_0) =D_r\R^2\ups(\pp,t\,|\,\pp_0, t_0)\,,
\end{equation}
where $\R\equiv(\pp\times\nabla_{\pp})$ is the intrinsic angular momentum 
differential operator. Eq.\eqref{FP_p} describes nothing but a diffusion 
process on the unit sphere. This problem is well-known when studying, 
e.g., dielectric relaxation in polar liquids 
\cite{debye1929,fatuzzo1967,zwanzig1970,berne1975}. In spherical coordinates, 
\eqref{FP_p} becomes
\begin{align}\label{FP_spher}
 \frac{1}{D_r}\frac{\partial}{\partial t}\ups(\Omega,t\,|\,\Omega_0,t_0) &=
 \left[\frac{1}{\sin \vt}\frac{\partial}{\partial 
\vt}\left(\sin \vt\frac{\partial}{\partial 
\vt}\right)\right.\notag\\
&\phantom{=}\left.
+\frac{1}{\sin^2\vt}\frac{\partial^2}{\partial\vp^2}\right]
\ups(\Omega,t\,|\,\Omega_0,t_0),
\end{align}
where we have defined $\Omega\equiv(\vt,\vp)$. 

Assuming that $\ups$ and its derivatives are continuous on the sphere \cite{arfken2013}, 
we expand the probability distribution function $\ups$ in spherical harmonics
\begin{equation}\label{solexp}
\ups(\Omega,t\,|\,\Omega_{0},t_0)=\sum_{l=0}^{\infty}\sum_{m=-l}^{l}A_{lm}
(t\,|\,\Omega_0,t_0)Y_{lm}(\Omega)\,,
\end{equation}
where $Y_{lm}$ are the spherical harmonics and $A_{lm}$ are coefficients 
encoding the initial condition.
We also recall that spherical harmonics are 
eigenvectors of the operator 
$\R^2$ (in spherical coordinates), namely that
\begin{equation}\label{eigenvectors}
 \R^2 Y_{lm}=-l(l+1)Y_{lm}\,.
\end{equation}
Inserting \eqref{solexp} in \eqref{FP_spher} and using \eqref{eigenvectors} 
we obtain
\begin{align}\label{prev_orth}
 \sum_{l,m}\frac{\partial}{\partial 
t}&A_{lm}(t\,|\,\Omega_0,t_0)Y_{lm}(\Omega) =\notag\\
&-D_r\sum_{l,m}
l(l+1)A_{lm}(t\,|\,\Omega_0,t_0)Y_{lm}(\Omega).
\end{align}
Multiplying both sides of \eqref{prev_orth} by $Y_{l'm'}^{\ast}(\Omega)$, integrating 
over solid angle and using the orthogonality property, 
$\int d\Omega\, 
Y_{l'm'}^{*}(\Omega)Y_{lm}(\Omega)=\delta_{m,m'}\delta_{l,l'}$, yields
\begin{equation}
 \frac{\partial}{\partial 
t}A_{lm}(t\,|\,\Omega_0,t_0)=-D_rl(l+1)A_{lm}(t\,|\,\Omega_0,t_0)\,,
\end{equation}
which has the solution 
\begin{equation}
 A_{lm}(t\,|\,\Omega_0,t_0)=e^{-D_rl(l+1)(t-t_0)}a_{lm}(\Omega_0)\,,
\end{equation}
where the $a_{lm}$ are a new set of coefficients. 
The probability distribution is thus given by
\begin{equation} 
\ups(\Omega,t\,|\,\Omega_0,t_0)=\sum_{l,m}e^{-D_rl(l+1)(t-t_0)}a_{lm}(\Omega_0)Y_{lm
}(\Omega).
\end{equation}
The initial condition,
\begin{equation}
 \ups(\Omega,t_0\,|\,\Omega_0,t_0)=\delta(\Omega-\Omega_0)\,,
\end{equation}
together with the completeness relation of the spherical harmonics,
\begin{equation}
\delta(\Omega-\Omega_0)=\sum_{l=0}^{\infty}\sum_{m=-l}^{l}Y_{lm}(\Omega)Y_{l
m}^{\ast}(\Omega_0),
\end{equation}
allows the missing coefficients to be identified,
\begin{equation}
 a_{lm}(\Omega_0)=Y_{lm}^{\ast}(\Omega_0)\,.
\end{equation}
The conditional probability distribution is now fully determined as
\begin{equation}\label{sol_pdf}
\ups(\Omega,t\,|\,\Omega_{0},t_0)=\sum_{l=0}^{\infty}\sum_{m=-l}^{l}e^{
-D_rl(l+1)(t-t_0)}Y_{lm}^{\ast}(\Omega_0)Y_{lm}(\Omega).
\end{equation}
As $t\to\infty$ only the terms with $l=0$ survive. 
The steady-state distribution function is thus given by
\begin{equation}\label{eq_distr}
 \ups_{eq}(\Omega)=\lim_{t\to\infty} 
\ups(\Omega,t\,|\,\Omega_{0},t_0)
=(4\pi)^{-1}.
\end{equation}

The conditional and equilibrium distributions, \eqref{sol_pdf} and \eqref{eq_distr}, 
respectively, can be used to coarse-grain the exact 
Langevin equations \eqref{full_langevin} and \eqref{langevin_orientation}. 
The approach taken is to consider the orientation vector $\pp_i(t)$ attached to 
particle $i$ as a stochastic variable and to provide its full statistical 
characterization. 
In spherical coordinates the orientation vector is given explicitly by 
\begin{align} 
\pp(t)&=(p_x(t),p_y(t),p_z(t))^{T}\notag\\
&=(\cos\vp(t)\sin\vt(t),
\sin\vp(t)\sin\vt(t),\cos\vt(t))^ { T },
\end{align}
where $\vp$ and $\vt$ are the azimuthal and polar angles respectively. Using 
\eqref{eq_distr} 
we have that
\begin{align}
 \langle p_z(t)\rangle &=\int d\Omega\, \ups_{eq}(\Omega)\cos\vt=0, 
\end{align}
together with analagous results for the $x$ and $y$ components
\begin{equation}
 \langle p_x(t)\rangle = 0 = \langle p_y(t)\rangle\,.
\end{equation}
Defining the new stochastic variable by $\X_i(t)\equiv v_0\,\pp_i(t)$, its first 
moment is thus given by
\begin{equation}\label{correlations}
 \langle\X_i(t)\rangle=v_0\langle\pp_i(t)\rangle=\boldsymbol{0}.
\end{equation}

Calculation of the equilibrium correlation matrix requires the conditional 
probability distribution function given by \eqref{sol_pdf}. For example, for the $zz$ 
component, we obtain
\begin{align}
\langle p_z(t)p_z(t_0)\rangle &=
\!\!\!\int\!\!d\Omega\!\!\int\!\! 
d\Omega_0 \cos\vt \cos\vt_0 
\ups(\Omega,t\,|\,\Omega_0,t_0)\ups_{eq}(\Omega_0)\notag\\
&=
\frac{1}{3}\int d\Omega \int
d\Omega_0\sum_{l,m}e^{-D_rl(l+1)(t-t_0)}\times\notag\\
&\phantom{=}
\times Y_{10}^{\ast}(\Omega)Y_{lm}(\Omega)Y_{10}
(\Omega_0)Y^{\ast}_{lm}(\Omega_0)\notag\\
&=
\frac{1}{3}e^{-2D_r|t-t_0|},
\end{align}
where we have expressed the cosine functions in terms of spherical harmonics, 
$Y_{10}=\sqrt{3/(4\pi)}\cos\vt=Y_{10}^{\ast}$, and used the orthogonality 
property. 
Calculations for the $xx$ and $yy$ 
components are performed in the same spirit. 
We thus obtain
\begin{equation}
\langle p_x(t)p_x(t_0)\rangle=\frac{1}{3}e^{-2D_r|t-t_0|}=\langle 
p_y(t)p_y(t_0)\rangle,
\end{equation}
whereas off-diagonal components of the correlation matrix are all zero. 
We can thus conclude that 
\begin{equation}\label{correlations2}
 \langle\X_i(t)\X_j(t')\rangle=v_0^2\langle\pp_i(t)\pp_j(t')\rangle=
 \frac{v_0^2}{3}e^{-2D_r|t-t'|}\boldsymbol{1}\delta_{ij}.
\end{equation}
It has been shown \cite{ghosh2012} that the probability distribution 
function \eqref{sol_pdf} can be well approximated by an expression which 
generalizes the planar Gaussian function to the sphere. 
The new noise function $\X_i(t)$ is thus approximately Gaussian 
distributed with zero mean and exponentially decaying correlations. The 
coarse-grained Langevin equation \eqref{effective_langevin} thus describes 
a stochastic process with additive colored-noise.

\section{Approximate Fokker-Planck equation} \label{appendix2}
To derive from \eqref{effective_langevin} an approximate Fokker-Planck equation we 
apply the functional calculus methods of Fox \cite{fox1986}.  
We address the one-dimensional case before generalizing to higher dimension. 
Consider the stochastic differential equation
\begin{equation}\label{process}
 \dot{x}(t)= F(x) + g(x)\chi(t)\,,
\end{equation}
where $F(x)$ and $g(x)$ may be nonlinear functions in $x$. If $g(x)=1$, the 
process is then called {\it additive}, otherwise it is called 
{\it multiplicative}. The noise function $\chi(t)$ is by definition Gaussian distributed 
with zero mean. Its second moment determines whether it is a 
{\it white} or {\it colored} noise. As we are interested here in the case of 
additive colored noise we set $g(x)=1$.

In the framework of functional calculus, the Gaussian nature of $\chi(t)$ is expressed by the 
following probability distribution functional,
\begin{equation}\label{pdf}
 P[\chi]=N e^{-\frac{1}{2}\int ds\int ds' \chi(s)\chi(s')K(s-s')}\,,
\end{equation}
where the function $K$ is the inverse of the $\chi$ correlation function and 
the normalization constant is expressed by a path integral over $\chi$,
\begin{equation}\label{N}
 N^{-1}=\int D[\chi] e^{-\frac{1}{2}\int ds\int ds' 
\chi(s)\chi(s')K(s-s')}\,.
\end{equation}
The first and second moments of $\chi$ are given by
\begin{align}
 \langle\chi(t)\rangle &= 0\,,\label{m1}\\
 \langle\chi(t)\chi(s)\rangle&=C(t-s)\,.
\label{m2}
\end{align}
Recalling that the functional derivative may be defined according to
\begin{equation}\label{f_deriv}
 \frac{\delta 
I[\phi]}{\delta\phi(t')}= 
\left.\frac{d}{d\lambda}I[\phi(t)+\lambda\delta(t-t')]\right|_{\lambda=0}\,,
\end{equation}
we now derive two useful identities. The first concerns the functional 
derivative of the probability distribution functional,
\begin{align}
 \frac{\delta P[\chi]}{\delta \chi(t)} &= \frac{\delta N}{\delta 
\chi(t)}e^{-\frac{1}{2}\int ds\int ds'\chi(s)\chi(s')K(s-s')}\notag\\
&\phantom{=} + 
N\frac{\delta}{\delta \chi(t)}e^{-\frac{1}{2}\int ds\int 
ds'\chi(s)\chi(s')K(s-s')}\notag\\
&=
-P[\chi]\int ds K(t-s)\chi(s)\,,
\label{dp/df}
\end{align}
where, using \eqref{f_deriv} and \eqref{m1}, it can be easily shown that $\delta 
N/\delta \chi(t)=0$. The second identity demonstrates the inverse relation between the 
functions $K$ and $C$. 
The second functional derivative of $P[\chi]$ yields
\begin{align}\label{2nd_deriv}
 \frac{\delta^2P[\chi]}{\delta\chi(t')\delta\chi(t)}&=
 P[\chi]\left\{\!\int\!\!ds'\!\!\!\int\!\!ds 
K(t'-s')K(t-s)\chi(s')\chi(s)\right.\notag\\
 &\phantom{=}
 \left.\phantom{\int}\!\!- K(t-t')\right\}\,,
\end{align}
where use of \eqref{dp/df} has been made. Using \eqref{2nd_deriv} and \eqref{m2} 
together with the normalization $\int D[\chi]P[\chi]=1$, leads to
\begin{align}
0 &= \int D[\chi]\frac{\delta^2P[\chi]}{\delta\chi(t')\delta\chi(t)}\notag\\
&=\int\!\!ds'K(t'-s')\int \!\!ds K(t-s)C(s-s')-K(t-t')\,,
\end{align}
which implies that
\begin{equation}\label{delta}
 \int \!\!ds K(t-s)C(s-s')=\delta(t-s')\,. 
\end{equation}

The solution to the stochastic process described by \eqref{process}, namely the 
probability distribution functional for $x(t)$, is given by the formal expression
\begin{equation}\label{sol}
 P(y,t) = \int D[\chi]P[\chi]\delta(y-x(t))\,.
\end{equation}
Taking the time derivative of \eqref{sol} yields
\begin{align}\label{towardsFP}
 \frac{\partial}{\partial t}P(y,t) &= 
-\frac{\partial}{\partial y}[F(y)P(y,t)]\notag\\
&\phantom{=}
-\frac{\partial}{\partial y}\int D[\chi]\delta(y-x(t))P[\chi]\,\chi(t)\,.
\end{align}
The product $P[\chi]\,\chi(t)$ appearing in the second term can be rewritten in 
the following way
\begin{align}\label{PX}
 P[\chi]\,\chi(t) &= P[\chi]\int ds \delta(t-s)\chi(s)\notag\\
 &=
 -\int ds' C(t-s')\frac{\delta P[\chi]}{\delta \chi(s')}\,,
\end{align}
where we have used \eqref{delta} and \eqref{dp/df}. Inserting 
\eqref{PX} back into the second term of \eqref{towardsFP} and integrating by 
parts gives us
\begin{align}\label{45}
 &\int D[\chi]\delta(y-x(t))P[\chi]\chi(t)\notag\\
 &=
 -\int\!\! ds' C(t-s')\!\!\int\!\! D[\chi]\left(\frac{\partial}{\partial 
y}\delta(y-x(t)) \right)\frac{\delta x(t)}{\delta \chi(s')}P[\chi], 
\end{align}
which serves as the exact starting point for Fox's approximation scheme \cite{fox1986}. 

In order to progress further we need to calculate $\delta x(t)/\delta 
\chi(s')$. Applying the functional derivative with respect to $\chi(t')$ on 
\eqref{process} yields a first-order differential equation,
\begin{align}
  \frac{d}{dt}\frac{\delta x(t)}{\delta \chi(t')} &=
  \frac{\delta \dot{x}(t)}{\delta\chi(t')}
  =F'(x)\frac{\delta x(t)}{\delta \chi(t')} +\delta(t-t')\,,
\end{align}
the solution of which is 
\begin{align}\label{dx/dX}
 \frac{\delta x(t)}{\delta \chi(s')} &= \int_{0}^{t}\!\!\!ds\, 
e^{\int_{s}^{t}d\tilde{s}F'(x(\tilde{s}))}\delta(s-s')\notag\\
&=e^{\int_{s'}^{t}ds F'(x(s))}\Theta(t-s')\,,
\end{align}
where $\Theta$ is the Heaviside step function, which we define here as follows
\begin{displaymath}
\Theta(t-s') = \left\{ \begin{array}{ll}
1, & t>s'\\
\frac{1}{2}, & t=s'\\
0, & t<s'\,.
\end{array} \right.
\end{displaymath}
Using \eqref{dx/dX} in \eqref{45} we can rewrite \eqref{towardsFP} in an alternative 
form
\begin{align}\label{46}
 &\frac{\partial}{\partial t}P(y,t) = 
 -\frac{\partial}{\partial y}[F(y)P(y,t)]\notag\\
 &
 +\frac{\partial^2}{\partial 
y^2}\!\!\left\{\!\int_{0}^{t}\!\!\!\!ds'C(t-s')\!\!\!\int\!\!\! 
D[\chi]P[\chi]e^{\int_{s'}^{t}ds F'(x(s))}\delta(y-x(t))\!\!\right\},
\end{align}
which already begins to resemble a Fokker-Planck-type equation. 
However, because of the 
non-Markovian nature of $\int_{s'}^{t}ds F'(x(s))$ appearing in the exponential of 
\eqref{46}, it is apparent that a reduction of this term to an expression 
containing $P(y,t)$ is not possible. An approximation is required. 

The colored noise of interest here is characterized by an 
exponentially decaying correlation function \eqref{correlations}. 
In the literature on non-Markovian processes the time-correlation functions 
are generally notated as follows
\begin{equation}\label{correl}
 C(t-s)=\frac{D}{\tau}e^{-\frac{|t-s|}{\tau}}\,,
\end{equation}
with a diffusion coefficient $D$ and a correlation time $\tau$. 
In order to retain some coherence with the existing literature we will here 
employ the standard 
notation of \eqref{correl} and only use the relation of the parameters in \eqref{correl} 
to those of \eqref{correlations2} at the end of the calculation. 

Returning to \eqref{46}, we 
first perform a change of variable, $t'\equiv t-s'$, in the time-integral,
\begin{equation}\label{nonmarkov}
 \int_{0}^{t}ds'C(t-s')e^{\int_{s'}^{t}ds F'(x(s))}
 =\int_{0}^{t}dt'C(t')e^{\int_{t-t'}^{t}ds F'(x(s))},
\end{equation}
and then expand the time integral over $F'$ in terms of $t'$, 
\begin{equation}\label{approx}
 \int_{t-t'}^{t}ds F'(x(s))\approx F'(x(t))t'-F''(x(t))\dot{x}(t)\frac{t'^2}{2}.
\end{equation}
Neglecting the $t'^2$ term in \eqref{approx} enables the integral in \eqref{nonmarkov} 
to be evaluated
\begin{align}\label{large_t}
 &\int_{0}^{t}ds'C(t-s')e^{\int_{s'}^{t}ds F'(x(s))} \approx 
 \int_{0}^{t}dt'C(t')e^{F'(x(t))t'}\notag\\
 &=\frac{D}{\tau}\int_{0}^{t}dt'e^{-t'(-F'(x(t))+\frac{1}{\tau})}
 \approx \frac{D}{1-\tau F'(x(t))}\,,
\end{align}
where we used \eqref{correl}, and the second approximation results from 
assuming a sufficiently large $t$. We can finally put \eqref{large_t} back into 
\eqref{46} to obtain an approximate Fokker-Planck equation,
\begin{align}\label{approxFP}
 \frac{\partial}{\partial t}P(y,t) &\!=\! 
 -\frac{\partial}{\partial y}[F(y)P(y,t)]+\!D\frac{\partial^2}{\partial 
y^2}\!\!\left(\!\frac{1}{1-\tau F'(y)}P(y,t)\!\right).
\end{align}
This is Fox's result for the approximate Fokker-Planck equation corresponding to 
the non-Markovian process \eqref{process}. Eq.\eqref{approxFP} 
implicity defines a Markovian process, which approximates the non-Markovian 
process of physical interest. However, the question of whether this represents 
the best approximation remains a subject of debate. 
We note that equation \eqref{approxFP} has also been derived by Grigolini {\it et al.} 
\cite{faetti1988} using alternative methods which do not make any assumptions of a short correlation 
time. 

The one-dimensional Fokker-Planck equation \eqref{approxFP} can be generalized without 
much difficulty to describe a three-dimensional system of $N$ particles. 
The dynamics of interest is described by the stochastic 
equation \eqref{effective_langevin}. 
We now adapt the standard notation used above to that employed in the main text, 
namely $P(y,t)\rightarrow \Psi(\boldsymbol{r}^N,t)$, 
$\tau\rightarrow\tau_p=1/(2D_r)$ and $D\rightarrow v_0^2/3$, and recall that 
$D_a=v_0^2/(6D_r)$ and $\zeta^{-1}=\beta D_t$ for the friction coefficient in 
\eqref{effective_langevin}. 
Making the appropriate replacements enables us to write 
the three-dimensional generalization of \eqref{approxFP} 
\begin{align}\label{1st_smol}
 \frac{\partial}{\partial t}&\Psi(\boldsymbol{r}^N,t)=
 -\sum_{i=1}^{N}\nabla_i\cdot
 D_t\left[\beta 
\boldsymbol{F}_i(\boldsymbol{r}^N)-\nabla_{i}\right]\Psi(\boldsymbol{r}^N,
t)\notag\\
 &
-\sum_{i=1}^{N}\nabla_i\cdot \left[-D_a\nabla_{i}\left(\frac{1}{1-\frac{
D_0\nabla_i\cdot\beta\boldsymbol{F}_i(\boldsymbol{r}^N)}{2D_r}}\Psi(\boldsymbol{
r}^N,t) \right)\right].
\end{align}
A simple rearrangement of terms in \eqref{1st_smol} leads directly to  
equations \eqref{smol_eq}-\eqref{force_i}
in the main text.


\section{Integral equation theory}\label{integralequation} 
To calculate the steady-state radial distribution function, $g(r)$,  
from the effective pair potential \eqref{potential_pair} we employ an equilibrium 
liquid state integral equation developed by Madden and Rice \cite{madden1980}. 
This soft mean-spherical approximation (SMSA) exploits the Weeks-Chandler-Anderson 
splitting of the pair potential \cite{WCA} into attractive and repulsive contributions,  
$u(r)=u_{\rm rep}(r) + u_{\rm att}(r)$, where the repulsive part is given by
\begin{align}
u_{\rm rep}(r)=
\begin{cases}
u(r)-u(r_{\rm min}) & \;\;\;\;r < r_{\rm min}
\\
0 &\;\;\;\; r > r_{\rm min}
\end{cases}
\end{align}
and the attractive part is given by
\begin{align}
u_{\rm att}(r)=
\begin{cases}
u(r) & \;\;\;\;r > r_{\rm min}
\\
u(r_{\rm min}) &\;\;\;\; r < r_{\rm min}
\end{cases}
\end{align}
where $r_{\rm min}$ is the position of the potential minimum. 
The total correlation function, $h(r)=g(r)-1$, is related to the shorter 
range direct correlation function, 
$c(r)$, by the Ornstein-Zernike equation \cite{hansen_mcdonald1986}
\begin{align}
h(r)=c(r) + \rho_b\int d{\rr}' h(|\rr - \rr'|)c(r').
\end{align}
The SMSA approximation is given by the closure relation 
\begin{align}\label{SMSA}
c(r)=(1 - e^{\beta u_{\rm rep}(r)})g(r) - \beta u_{\rm att}(r). 
\end{align} 
For the Lennard-Jones potential the closure relation \eqref{SMSA} has been shown 
to provide results for $g(r)$ which are superior to both Percus-Yevick (PY) and 
Hypernetted Chain (HNC) theories \cite{madden1980}. 
Moreover, the SMSA theory predicts a true spinodal line 
in the parameter space, namely a locus of points for which the static structure 
factor, $S(k)=(1-\rho_b\tilde{c}(k))^{-1}$, diverges at vanishing wavevector. 
This behaviour is a consequence of the assumed asymptotic form of the direct correlation 
function, $c(r)\sim -\beta u_{\rm att}(r)$. 
Other standard integral equation theories, such as PY and HNC, do not exhibit a 
complete spinodal line, but rather a region within which the theory breaks down 
(`no solutions region')
\cite{braderIJTP}.

\section{Brownian Dynamics Simulations} \label{simulations}
In order to benchmark our theoretical predictions we perform Brownian dynamics 
simulations of $N$ particles, randomly initialized without overlap. 
The system is confined to a periodic cubic box, the size of which 
is determined by the number density according to 
$L^3=N/\rho_b$, where $L$ is the side length. 
The Langevin equations of motion, \eqref{full_langevin} and \eqref{langevin_orientation},  
are integrated via a standard Brownian 
dynamics scheme \cite{allen_tildesley} with a constant time step of $\delta t / 
\tau_B = 10^{-5}$.\!
Both the translational and rotational noise are 
Gaussian random variables with a standard deviation of $\sigma_t\!=\!(2 D_0 T)^{\frac{1}{2}}$ 
and  $\sigma_r\!=\!(2 D_r T)^{\frac{1}{2}}$, respectively. 

For the soft repulsive potential to be considered in this work, $\beta u(r)=r^{-12}$, 
we employ $N\!=\!2000$ particles. 
The potential is truncated and shifted at $r_{\rm cut} / d = 2$. 
To provide good statistics for the static quantities the 
simulations are carried out for $10^6$ time steps, sampling every 1000 
steps, which is equivalent to a total run time of $t_{\rm tot} / \tau_B = 10$ and a 
sampling rate of $\tau_B / t_{sample} = 100$.
For the second system we will consider, the Lennard-Jones system, 
$\beta u(r)\!=\!4\epsilon(r^{-12}\! -r^{-6})$,
we simulate a larger 
system of 5000 particles. 
The integration time of the 
equations of motion is the same as in the repulsive system, as is the 
cut-off radius. In this case, the runtime is $10^7$ and the particle positions are sampled 
every $10^4$ steps. 

\end{document}